\documentclass[preprint2]{aastex701}

\usepackage{amssymb}
\usepackage{lipsum}
\usepackage{bm}
\usepackage{amsmath}
\usepackage{wasysym}
\usepackage{cancel}
\usepackage{xspace}
\usepackage{comment}

\newcommand{\pdf}{\mathrm{P}}

\newcommand{\kep}{Kepler-167\,e\@\xspace}
\newcommand{\jwst}{\textit{JWST}\@\xspace}
\newcommand{\kepler}{\textit{Kepler}\@\xspace}

\newcommand{\fortran}{\texttt{fortran}\@\xspace}
\newcommand{\multi}{\texttt{MultiNest}\@\xspace}

\newcommand{\ultra}{\texttt{UltraNest}\@\xspace}

\newcommand{\luna}{\texttt{LUNA}\@\xspace}

\newcommand{\pandexo}{\texttt{pandexo}\@\xspace}
\newcommand{\pandeia}{\texttt{pandeia}\@\xspace}

\newcommand{\matern}{{\tt M32}}
\newcommand{\gp}{{\tt SE}}
\newcommand{\expn}{{\tt exp}}
\newcommand{\qudr}{{\tt quad}}

\newcommand{\custom}{\texttt{custom}\@\xspace}
\newcommand{\eureka}{\texttt{Eureka!}\@\xspace}
\newcommand{\jedi}{\texttt{ExoTiC-JEDI}\@\xspace}
\newcommand{\kat}{\texttt{katahdin}\@\xspace}
\newcommand{\jwstpipe}{\texttt{jwst}\@\xspace}

\newcommand{\vstack}[2]{\shortstack[c]{#1\\[1pt]#2}}

\begin{document}


\title{A JWST Transit of a Jupiter Analog: \\II. A Search for Exomoons}

\author[orcid=0000-0002-4365-7366,sname='Kipping']{David Kipping}
\affiliation{Columbia University, 550 W 120th Street, New York NY 10027, USA}
\email[show]{dkipping@astro.columbia.edu}

\author[orcid=0000-0002-9544-0118]{Ben Cassese}
\affiliation{Columbia University, 550 W 120th Street, New York NY 10027, USA}
\email{b.c.cassese@columbia.edu}

\author[0000-0001-6516-4493]{Quentin Changeat}
\affiliation{Kapteyn Institute, University of Groningen, 9747 AD Groningen, NL}
\email{qchangeat@stsci.edu}

\author[0000-0003-4755-584X]{Daniel Yahalomi}
\affiliation{Department of Physics and Kavli Institute for Astrophysics and Space Research, Massachusetts Institute of Technology, Cambridge, MA 02139, USA}
\email{yahalomi@mit.edu}

\author[0000-0003-2331-5606]{Alex Teachey}
\affiliation{Academia Sinica Institute of Astronomy \& Astrophysics, Roosevelt Rd, Taipei 10617, Taiwan (R.O.C.)}
\email{amteachey@asiaa.sinica.edu.tw}

\author[0000-0002-5494-3237]{Billy Edwards}
\affiliation{SRON, Netherlands Institute for Space Research, Niels Bohrweg 4, NL-2333 CA, Leiden, The Netherlands}
\email{B.Edwards@sron.nl}

\begin{abstract}
We present a search for exomoons around the Jupiter-like exoplanet Kepler-167e using a NIRSpec JWST transit. Our 60\,hour time series clearly reveals the enormous impact of long-term trends in NIRSpec data, specifically a gradual flux drift occurring over each of six 10\,hour exposures. We weighed the evidence for exomoons by comparing a planet-only model with a planet+moon model for a grid of twelve different analysis choices. Our grid was comprised of three different reduction pipelines and four different models for the exposure-long trends - two using linear models and two using Gaussian processes. Seven grid realizations indicate a strong exomoon detection, typically favoring a Roche-skimming orbit roughly 10\% the size of the planet. We find that the only likely real astrophysical feature driving these fits is a syzygy-like event occurring almost exactly mid-transit, which is highly ambiguous with a spot-crossing event. Indeed, we show that a spot of the necessary size is compatible with the earlier \kepler\ data. Ironically, the fact that JWST is so superior to \kepler\ means that our fits are effectively driven by a single transit - a regime in which exomoons have enormous freedom to explain non-Gaussian behavior. We thus strongly urge the next transit be observed in October 2027 to break these degeneracies. Our pilot study to seek transiting exomoons with JWST reveals the profound impact exposure-long trends exert - a cautionary tale for future analyzes of this data - as well as the need for a deeper understanding of this systematic's cause and modeling best-practices.
\end{abstract}



\textcolor{white}{blank}\\
\textcolor{white}{blank}\\
\textcolor{white}{blank}

\section{Introduction}
\label{sec:intro}

JWST offers the opportunity to capture photometric transits at an unprecedented level of precision. Although most transit studies with JWST focus on its spectroscopic capabilities to probe atmospheres, the summed white light curves allow us to constrain various other effects like oblateness, rings and exomoons via transits.

Searches for transiting exomoons with \kepler\ have indicated that Earth-sized (or larger) moons are uncommon \citep{hek6} for planets between 0.1-1\,AU. At wider orbits, two compelling candidates have been reported around long-period gas giants, Kepler-1625\,b \citep{teachey:2018} and and Kepler-1708\,b \citep{k1708}. In both cases, the detection papers caution the need for follow-up observations given the limited data available. \citet{heller:2024} challenged the validity of these candidates in a re-analysis of the original data, although it was later demonstrated that their fits had simply missed the global minimum solution, amongst other issues \citep{kipping:2025}.

Besides these two, there have been numerous other claims for exomoons, both with and without the transit technique. For example, \citet{bennett:2014} reported a microlensing event with two degenerate solutions, one of which corresponded to a moon around a free floating planet. A decade later, Keck follow-up confirmed it was in fact the non-exomoon solution \citep{terry:2025}. In another case, \citet{cabrera:2014} identified a moon-like dip near the transit of the long-period gas giant Kepler-90g. However, later analysis of the pixel-level photometry revealed the dip to be likely a sudden pixel drop out effect \citep{kipping:2015}. In a third example, \citet{fox:2021} found multiple \kepler\ planets exhibiting transit timing variations (TTVs) consistent with exomoons, but the robustness of these signals was challenged in \citet{kipping:2020}. Other notable claims include ring gaps carved by moons for J1407 \citep{kenworthy:2015}, disintegrated moons to explains the light curve of Boyajian's Star \citep{martinez:2019}, a possibly moon-forming disk imaged around PDS-70c \citep{benistry:2021} and sodium emission interpreted as possible volcanic outgassing of a moon around the hot-Jupiter WASP-49\,A\,b \citep{oza:2024}.

There is certainly an opportunity for JWST to study these elusive objects given its spectacular precision and perhaps finally deliver an unambiguous detection. Long-period, massive planets are the natural hunting ground for moons, since they have larger Hill spheres and have experienced less dramatic inward migration \citep{barnes:2002,domingos:2006,namouni:2010}. Planets in multi-planet systems yet retaining circular orbits are also attractive, indicating a lack of destructive scattering history \citep{gong:2013}. Kepler-167e satisfies all of these criteria and is arguably the most Jupiter-like transiting exoplanet ever discovered \citep{kipping:2016}. With an orbital period of approximately three years though, opportunities to observe the planet are few and far between. Thus, the fact a transit was expected in Cycle 3 of JWST sparked our interest to try to not only catch an elusive transit of a Jupiter-analog, but conduct the first-ever transit search for exomoons using this facility. Our goal was to better understand what JWST is capable of in this domain, and inform future surveys for exomoons with this outstanding facility. Although this paper focusses on the exomoon hunt, a companion paper \citep{paper1} instead explores our sensitivity to planetary oblateness using the same data set.

\section{Observations}
\label{sec:observations}

\subsection{Data Collection}
\label{sub:collection}

Our dataset comprises a 59.78\,hour time series obtained with the NIRSpec instrument on board \textit{JWST}, centered on the predicted transit of \kep\ on 2024~October~25. Observations employed the Bright Object Time Series configuration together with the Prism/Clear disperser and filter, providing continuous spectroscopy over 0.6–5.3,$\mu$m at a typical resolving power of $R\approx100$. We adopted the NRSRAPID readout pattern, which records every frame individually (i.e., without onboard frame averaging), as is standard practice for time-series applications. The standard SUB512 subarray was used—this is a 32$\times$512 pixel region of the NRS1 detector that excludes reference pixels—with six groups per integration and a total of 134{,}208 integrations. Altogether, the cube comprises $n_{\mathrm{pix}}\times n_{\mathrm{groups}}\times n_{\mathrm{integrations}} = 1.319\times10^{10}$ individual flux samples. To our knowledge, this constitutes the longest uninterrupted time-series observation yet acquired by \textit{JWST}.

Because of the extraordinary duration and the finite data-handling capacity of the observatory, the observation was segmented into six exposures separated by short pauses. Between exposures, the detector was reset and briefly entered passive continuous-readout mode. Exposures~1–2 precede any planetary transit, exposures~3–5 capture the transit of \kep, and exposure~6 includes a transit of Kepler-167\,c. As illustrated in Figure~\ref{fig:example} and discussed in Section~\ref{sec:trendmodels}, each exposure exhibits its own gradually decreasing, nonlinear background trend.
To ensure the robustness of our findings, we produced and analyzed three independent data reductions, detailed below.

\subsection{\custom\ Reduction}
\label{sub:custompipeline}

Our first reduction adopts a philosophy of conservatism. It follows the standard initial reduction of the STScI pipeline but then asserts aggressive cuts to ensure the final white light curve behaves as close to independent Gaussian noise as possible. We simply dub this as our \custom\ reduction in what follows.

We began by running the \texttt{Detector1Pipeline} of \jwstpipe for our Stage 1 reduction, or conversion from uncalibrated 4D counts to a 3D data cube of rates. We used the default pipeline arguments aside from \texttt{clean\_flicker\_noise}, which we set to ``median'', and the ramp-fitting step, which we set to ``LIKELY'' in order to use the algorithms from \citet{brandt_ramp_fit_2024} and \citet{brandt_jump_detect_2024}. Recently, \citet{carter_use_brandt_2025} demonstrated that these algorithms offered superior performance for NIRISS/SOSS data. We then ran the \texttt{Spec2Pipeline} of \jwst to assign a WCS and wavelength mapping for our Stage 2 reduction.

We then considered dark frames from the Cycle 2 calibration program \#4456, CAL-NRS-202 Dark Monitor Subarray \citep{jwst_darks}. This program recorded 7 exposures of 5 integrations and 340 groups/integration using the same SUB512 subarray used in our observations. We used these to identify all pixels that exhibit either anomalously high variance or Random Telegraph Noise behavior. This phenomenon refers to a subset of pixels whose bias/reset levels randomly and spontaneously jump by several thousand data numbers. Since \jwstpipe does not fit for per-pixel, per-integration biases (unlike \kat, which does, and \jedi, which optionally sacrifices some data by using the first group as the bias frame), these pixels could in principle produce systematically incorrect rates. We removed 112 pixels in total suspected of either this behavior or high variance.


For each pixel, we create a time series and trim any fluxes more than 5\,$\sigma$ deviant from a 21-point moving median (where $\sigma$ is the pipeline error on each point). Pixel light curves are then summed into columns, since each column corresponds to a distinct wavelength, as well as being orthogonal to dispersion axis. For each column, we then define a background from rows 1-7 and 25-32. The summed background flux is subtracted from the aperture flux to create a time series for each column. We found that only columns 30 to 460 (inclusive) span the target flux and thus we focus on these 431 columns in what follows.


As part of this reduction, we perform a preliminary fit to the light curves to identify problematic columns. In total, we have 431 column light curves. Ideally, we would fit this entire array of data hierarchically but this immediately raises the spectre of a vast number of free parameters. A simple but effective trend model is to treat each exposure as a quadratic curve, but since we have 6 exposures per wavelength, then this is $6\times3=18$ trend parameters per wavelength, or $18 \times 431 = 7758$. On top of this, there are also the $431$ unique depths. Rather than attempting a full MCMC of $\mathcal{O}[10^4]$ free parameters, we employ a profile likelihood technique (discussed later in Section~\ref{sub:pl}). The global parameters, like impact parameter and stellar density, are treated as typical MCMC terms but the transit depths are selected by maximizing the likelihood function at any given MCMC position using the Golden Section algorithm \citep{press:2007}. As the Golden Section process makes guesses for the depths, the trend parameters are set by solving the normal equation. In this way, we greatly reduce the dimensionality but maintain some of the benefits of an MCMC approach.
Column-specific quadratic limb darkening coefficients were generated using the ExoTIC-LD package \citep{grant:2024}. Each coefficient was generated by linearly interpolating between stellar atmosphere models in the Kurucz grid that best match the fiducial $T_{\mathrm{eff}}$, $\log g$, and metallicity from \citet{chachan:2022}. These models were then integrated over the very narrow wavelength range defining each column, and the resulting intensity profile was approximated with a quadratic.
Profile likelihood techniques can lead to narrower posteriors than truth, but in our application we are primarily interested in the maximum likelihood solution and so this is not important.

We take the maximum likelihood residuals from a planet-only fit to search for outlier column light curves. As a first test, we measured the RMS of the residuals as a function of column index and used an interpolated moving median (bandwidth 5) to identify RMS values which are more than two standard deviations deviant (where the standard deviation was found using the MAD). We did the same for other metrics, specifically the Allan Variance, the Ljung-Box test and the Durbin-Watson statistic - which all test for time-correlated noise structure in each column light curve. Finally, we tracked the $\chi^2$ of each column's planet-fit and trimmed any 5\,$\sigma$ outliers similar to how we did so RMS. These tests identified 106 columns as being of insufficient quality, which were thus rejected.


We next considered whether anomalies in the spectral trace could be used to flag outlier columns. For each column, we measured the centroid position with as a simple flux weighted average (after removing know bad pixels) from row 7 to 24. For any given column, we removed any centroids in time more than 10\,$\sigma$ deviant from the median (where $\sigma$ is again computed from the MAD). We then median binned each column's centroids into 5-minute windows. We also calculated the trace width at each column index, defined as the FWHM using a spline interpolation (see \citealt{espinoza:2023}), which we binned the same way as the centroids. Using these, we constructed interpolated moving median functions of each as a function of column index. We then flagged columns for which the centroid was more than 0.05 pixels deviant, or the width was more than 0.1 pixels deviant. This trimmed another 34 columns as being of insufficient quality.


In total, then, 140 columns (out of 431 considered) were masked to create our final white light curve. This diminished the total flux from $5.2 \times 10^6$ counts per raw cadence to $4.8 \times 10^6$ (as measured using the median of exposures 1 and 2), and thus an effective 4\% decrease in photon SNR (which scales as square root counts). 

\subsection{\jedi}
\label{sub:exojedipipeline}

For our second reduction, we used one of the \jedi\ \citep{alderson_wasp_ers_2023} reductions from Paper I \citep{paper1}. Specifically, we used the version with destripping width 15 and aperture width 3. A fuller description of this pipeline and our application of it can be found in that article. 

\subsection{katahdin}
\label{sub:katahdinpipeline}

For our third reduction, we used the \kat pipeline, a new independent reduction framework developed over this analysis. An article describing katahdin is in prep, and a slightly expanded description can be found in Paper I \citep{paper1}. \kat serves the same function as Stage 1 of the \jwstpipe pipeline \citep{bushouse:2025}, meaning it converts uncalibrated ramp data into per-integration rate images. This manuscript diverges from Paper I in how it extracts a white light curve from these images. Instead of using the signal-to-noise thresholds described in that study, here we apply the same aperture/background removal procedure used in the custom reduction described in Section~\ref{sub:custompipeline}.

\subsection{Other Pipelines}

We briefly note that we also performed a \eureka\ reduction \citep{taylor:2022} but found that it gave consistently worst fit residuals, to such a dgeree many of our moon fits failed to even converge. The lower performance of \eureka\ versus \jedi\ and \kat\ is discussed in more depth in Paper I, but we do not consider is further here.

\subsection{Binning Procedure}
\label{sub:binning}

Each of the three pipelines described above created spectral light curves with 134,208 time stamps and varying numbers of wavelength channels. We summed along the spectral dimension and reshaped the data into five minute bins.

\subsection{Masking Planet c}

In all reductions, we mask the transit of Kepler-167c, using the ephemeris from \citet{kipping:2016}. This removes points between MJD 60610.448 and 60610.592.

\section{Trend Modeling}
\label{sec:trendmodels}

\subsection{Why Trends Matter}
\label{sub:trendsmatter}

Our observations include six exposures. In all of our reductions, we observe a gradual decrease in the white light curve intensity during each exposure, followed by a reset at transition points between each exposure.
Exposure-long trends have been observed across both ground-based testing \citep{rauscher:2014,zafar:2022}, but their physical origin remains unclear with the JWST User Documentation stating ``It is possible that this effect is due to some underlying detector level effect, and this issue is currently under investigation.'' \citep{jwst:2016}.

In our observations, the repeatability of these trends with respect to exposure number strongly indicates a detector effect, rather than an astrophysical origin. Like the visit-long trends observed in HST WFC3 data \citep{wakeford:2016}, the community typically employs heuristic models to correct for such trends, most commonly a simple linear function \citep{zafar:2023}.

In our case, each exposure spans approximately 10\,hours, which is comparable to the timescale we expect for exomoon-like dips. Accordingly, the accurate modeling of these exposure-long trends is critical in seeking exomoons - to such a degree we argue here it is the primary point of concern in any comparable search. Indeed, this mirrors the situation with WFC3 visit-long trends, where it has been previously demonstrated that varying the trend model between the most widely adopted community choices can lead to substantial differences in subsequent exomoon inference efforts \citep{teachey:2018,loose:2020}.

We therefore proceed carefully in our choice of trend model, testing several variants to assess their influence on our final results. We describe these different trend models in what follows.

\subsection{Terminology}
\label{sub:terminology}

A typical starting assumption in analyzing photometric time series is that the data, $y(t)$, is described by some astrophysical model, $f(t;\boldsymbol{\theta})$, scaled by some instrument model, $g(t;\boldsymbol{\phi})$ , hence one may write that

\begin{align}
y(t) = g(t;\boldsymbol{\phi}) f(t;\boldsymbol{\theta}) + \epsilon,
\end{align}

where $\boldsymbol{\theta}$ is a vector of the parameters describing the astrophysics (e.g. planet/moon parameters) and $\boldsymbol{\phi}$ is a vector of the parameters describing the instrument/trend terms. We initially assume the noise, $\epsilon$, is white and uncorrelated in time and thus our likelihood function would be the sum of independent Gaussians:

\begin{align}
\log\mathcal{L} &= - \frac{n}{2}\log(2\pi) - \tfrac{1}{2} \sum_{i=1}^n \log \sigma_i^2 - \frac{1}{2}\sum_{i=1}^n \Big(\frac{r_i}{\sigma_i}\Big)^2,
\label{eqn:loglike}
\end{align}

where $\mathbf{r}$ is the residuals vector given by $r_i = y_i - g(t_i;\boldsymbol{\phi}) f(t_i;\boldsymbol{\theta})$. Accordingly, such a treatment assumes $y_i \sim g(t_i;\boldsymbol{\phi}) f(t_i;\boldsymbol{\theta}) + \mathcal{N}(0,\sigma_i)$ i.e. the model plus some zero-mean Gaussian noise given by the formal errors.

Equipped with a likelihood function, Bayesian inference proceeds by evaluating the posterior probability of a large number of samples, to gradually converge upon the target distribution, where the posterior probability is given by

\begin{align}
\pdf(\boldsymbol{\theta},\boldsymbol{\phi}|\mathbf{y}) &\propto  \pdf(\mathbf{y}|\boldsymbol{\theta},\boldsymbol{\phi}) \pdf(\boldsymbol{\theta}) \pdf(\boldsymbol{\phi}),
\label{eqn:fullposterior}
\end{align}

where we have assumed the priors on $\boldsymbol{\theta}$ and $\boldsymbol{\phi}$ are independent of one another. Astrophysically speaking, we only care about $\boldsymbol{\theta}$, and thus one may marginalize over $\boldsymbol{\phi}$ to get the joint posterior of interest,

\begin{align}
\pdf(\boldsymbol{\theta}|\mathbf{y}) &\propto \pdf(\boldsymbol{\theta}) \underbrace{ \int  \pdf(\mathbf{y}|\boldsymbol{\theta},\boldsymbol{\phi}) \pdf(\boldsymbol{\phi}) }_{ \tilde{\mathcal{L}}(\mathbf{y}|\boldsymbol{\theta}) }.
\label{eqn:thetaposterior}
\end{align}

where we have labeled the integral as an effective likelihood function, $\tilde{\mathcal{L}}(\mathbf{y}|\boldsymbol{\theta})$, which will be useful later.

In the above scheme, many choices remain undefined, such as what is $f(t_i;\boldsymbol{\theta})$ and especially (in the context of Section~\ref{sub:trendsmatter}) what is $g(t_i;\boldsymbol{\phi})$? On top of this, the above only represents a starting point for a photometric time series analysis. Crucially, the assumption of white noise should be evaluated and modified as needed, as we will explore in what follows.

\subsection{Profile Likelihoods}
\label{sub:pl}

The first problem we run into in attempting to use the above scheme is the high dimensionality of the problem. If we adopted the typical approach of a first-order polynomial for each exposure, we would thus have two free parameters per exposure. But inspection of each exposure trend revealed that such a simplistic trend does not suffice and thus $g(t)$ will require at least three parameters per exposure. Since we have six exposures, this equates to 18 trend parameters in total (i.e. $M \equiv \mathrm{dim}(\boldsymbol{\phi})=18$). With 14 astrophysical parameters to describe the planet+moon hypothesis (i.e. $N \equiv \mathrm{dim}(\boldsymbol{\theta})=14$), the high dimensionality poses a challenge to Bayesian inference efforts.

A similar problem exists when fitting the much simpler \textit{Kepler} time series spanning a large number of transits. If each transit is assigned a unique flux offset term, the number of free parameters can quickly balloon. One solution, previously implemented in such cases (e.g. see \citealt{kundurthy:2011}), is to sample $\boldsymbol{\theta}$ and then at each such step set the nuisance parameters, $\boldsymbol{\phi}$, to be those which maximise the likelihood. This greatly reduces the dimensionality and still recovers the joint posterior of interest. This technique is sometimes called ``profile likelihood'' (PL) in the statistics community \citep{reid:2003,cousins:2024} and has recently been gaining traction in cosmology studies \citep{nygaard:2023,planck:2014,herold:2025}. A companion paper to this one pedagogically discusses our approach (Moynihan et al. 2025).

Moynihan et al. (2025) demonstrate that if the nuisance model is linear (i.e. we can construct a linear design matrix) and the errors are approximately homoscedastic, then the pseudo-posterior, $\pdf_{PL}(\boldsymbol{\theta}|\mathbf{y})$, that results from a PL approach is equivalent to the true one, $\pdf(\boldsymbol{\theta}|\mathbf{y})$. Further, so too is the marginal likelihood.

As numerical demonstration of the above, we used \multi\ \citep{feroz:2008,feroz:2009} to sample the astrophysical parameters, $\boldsymbol{\theta}$, and then minimize the remaining instrument/trend terms ($\boldsymbol{\phi}$) using PL. In this case, we adopt a quadratic trend for each exposure-long trend. For this test, we considered our simplest astrophysical model - a planetary transit of e only (masking c). With quadratic limb darkening, the model has $N=7$ astrophysical parameters, although orbital period was locked into to a Gaussian prior, so effectively 6 terms. These are: $p\equiv R_P/R_{\star}$, $b$, $\rho_{\star}$, $\tau$, $q_1$ and $q_2$.

We ran the fit twice with the same data and $\boldsymbol{\theta}$ prior: once using PL over the trend parameters and once using full marginalization that included 25 free parameters (close to the maximum dimensionality \multi\ can comfortably handle). We then compared the standard deviation of the marginalized posteriors for these six terms, giving ratios of $\{0.993,0.993,0.989,1.005,1.017,1.009\}$. Given that the pseudo-posterior includes 8634 posterior samples, sampling error means these are all indistinguishable from unity. It can therefore be seen that this example supports our claim that the pseudo-posterior accurately approximates the true one.

\subsection{Quadratic Trend}
\label{sub:quadtrend}

In considering different possible trends models, a quadratic model is an obvious starting place given its simplicity and literature precedence \citep{sarkar:2024}. Accordingly, we write that the $k^{\mathrm{th}}$ exposure is modeled as

\begin{align}
g_k(t) &= a_k + b_k (t-T_k) + c_k (t-T_k)^2,
\end{align}

where $a_k$, $b_k$ and $c_k$ are our free parameters and $T_k$ is the start of each exposure.

\begin{figure*}
\centering
\includegraphics[width=17.0 cm]{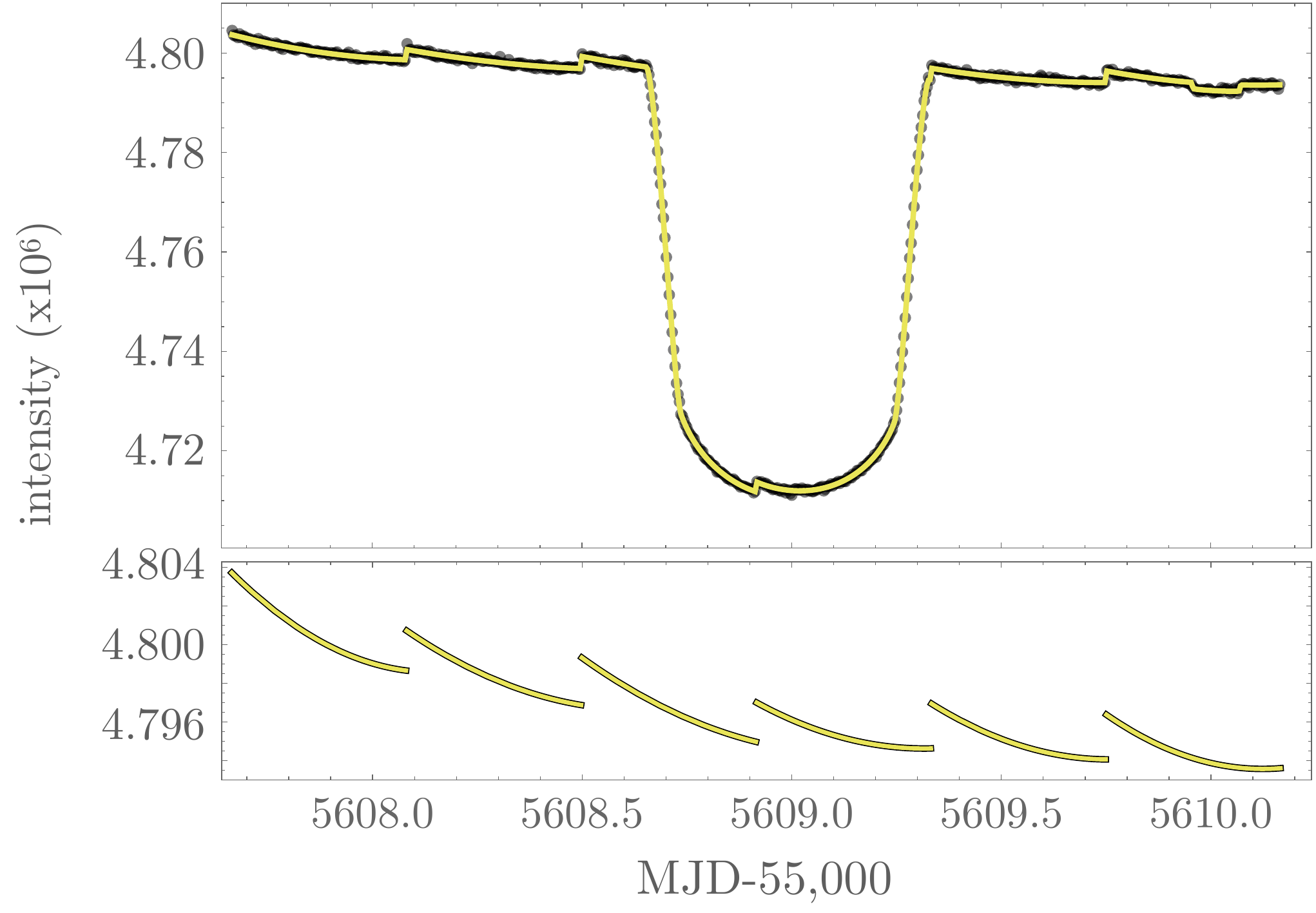}
\caption{\emph{
Broad view of the NIRSpec time series of Kepler-167e. Top shows the light curve from the \custom\ pipeline, overlaid in yellow with a simple planet-only model with quadratic limb darkening. Bottom zooms-in on the trends in isolation, highlighting the six distinct exposures of 10\,hours each. Note the transit of Kepler-167c also occurs in the last exposure, which we mask in what follows.
}
} 
\label{fig:example}
\end{figure*}

However, the model lacks any clear physical motivation and its simplicity allows for great flexibility, including exhibiting arguably unphysical behavior. For example, the trends can be non-monotonic, which for example occurs in the fourth and sixth exposures of \custom\ Figure~\ref{fig:example}.

Despite knowing the exposure-long trends are a detector-driven effect, the precise cause is not yet known \citep{jwst:2016}. However, a plausible cause is detector persistence, an issue known to affect NIRSpec \citep{rauscher:2014}. If this hypothesis were true, one would not expect the non-monotonic behaviour favored by the quadratic trends in Figure~\ref{fig:example}, and indeed that behaviour would then be merely an artifect of the limited freedom of the quadratic parameterization. Of particular concern is that such flips occurring around e's transit may interact with the exomoon model in complex ways, potentially leading to spurious moon signals as a result of the unrealistic flexibility of the quadratic model.

We thus conclude the quadratic model is a relevant benchmark model, but its excessive flexibility and tendency to produce ostensibly unphysical trends makes it a precarious choice. We thus proceed to consider more constrained models. Of course, the more constraints we place on our trend models, the worse the maximum likelihood (best fitting $\chi^2$) and associated RMS scores will be. This is an inevitable consequence of culling suspected over-fitting behavior, as we have here.

\subsection{Exponential Trend}
\label{sub:exptrend}

Aside from polynomial-based trends, exponential trends have also found widespread use (e.g. \citealt{zafar:2023}) and arguably have a better motivation in terms of detector physics \citep{agol:2010}. In particular, we adopted the same exponential trend model as that used by \citet{madhu:2025} which includes a linear and exponential term, such that the $k^{\mathrm{th}}$ exposure is modeled as

\begin{align}
g_k(t) &= a_k + b_k (t-T_k) + c_k \exp(-\lambda(t-T_k))
\end{align}

where $a_k$, $b_k$ and $c_k$ are our free parameters, $T_k$ is the start of the exposure and $\lambda$ is some pre-chosen drop-off rate. After some experimentation, we found that $1/\lambda=0.15$\,days provided consistently good fits for the different reductions and thus adopted this in what follows. We acknowledge the limitation of not marginalizing over $\lambda$ as part of our fits, but such a choice is necessary to keep the $g(t)$ strictly linear and thus satisfy the conditions for our profile likelihood implementation. We note that such a model still permits for non-monotonic behavior but we found this was less pervasive than when using quadratic trends.

We note that Paper I \citep{paper1} also uses an exponential trend but actually fits for $\lambda$, since it does not use the PL method where this is required. The PL method is not needed in Paper I since the oblateness model has fewer dimensions than the moon model and thus we found nested sampling could handle marginalizing out the trends. However, this was only possible in Paper I at the expense of dropping the linear time term, which we have here. These choices highlight some of the tradeoffs necessary in navigating high dimensional problems such as these.

\subsection{GP Trend}
\label{sub:gptrend}

Preliminary fits revealed that the exponential model is able to well explain the long term exposure trend, but there was evidence for smaller scale variability. In particular, our best fits to all reductions led to a reduced $\chi^2>1.3$ - indicating that our (thus far tacit) assumption of a strictly Gaussian likelihood function was questionable. As has been done in other recent similar analyzes (e.g. \citealt{espinoza:2025}), we thus instead tried modeling the trends with a zero-mean Gaussian Process (GP) by modifying our likelihood function \citep{gibson:2012}. The new log-likelihood function, $\log\mathcal{L}$, is given by

\begin{align}
\log\mathcal{L} &= - \tfrac{n}{2}\log2\pi - \tfrac{1}{2} \mathbf{r}^T \mathbf{K}^{-1} \mathbf{r} - \tfrac{1}{2} \log\mathrm{det}\mathbf{K} ,
\label{eqn:loglikeGP}
\end{align}

where $\mathbf{K}$ is a covariance matrix given by

\begin{align}
K_{ij} = k(t_i,t_j) + \sigma_i^2 \delta_{ij},
\label{eqn:kernel}
\end{align}

where $k(t_i,t_j)$ is our kernel function. We tried two different kernels, a squared exponential and a Mat\'ern-3/2. The former is characterized by an amplitude term, $\sigma_{SE}$, and a timescale $l_{SE}$:

\begin{align}
k_{SE}(t_i,t_j) &= \sigma_{SE}^2 \exp\Big(- \frac{(t_i-t_j)^2}{2l_{SE}^2} \Big)
\label{kern:SE}
\end{align}

We treated these two extra terms as global to the entire time series and included them as fully marginalized terms with a log-uniform prior on both. Thus, unlike the quadratic and exponential fits described previously, there is no profile likelihood approach here - just conventional sampling. We also tried a Matern-3/2 kernel:

\begin{align}
k_{M32}(t_i,t_j) =& \sigma_{M32}^2 \Big(1+\sqrt{3}\frac{|t_i-t_j|}{l_{M32}}\Big)\nonumber\\ 
\qquad& \times \exp\Big(-\sqrt{3}\frac{|t_i-t_j|}{l_{M32}}\Big)
\label{kern:M32}
\end{align}

Our implementation treats the global log-likelihood as a sum of six local log-likelihoods - one for each  exposure. In this way, the GP treats each exposure as a separate time series (which is why exposure jumps naturally are accounted for, e.g. see Figure~\ref{fig:m32custom}), but with shared hyper-parameters. We also experimented with Matern-1/2, quasi-periodic and simple harmonic oscillator kernels but found no substantial improvements.

Because the GP is defined to have zero mean, the model must be additive rather than multiplicative. However, since our nominal astrophysical model is a unit-normal light curve model, and the observed fluxes are $\gg1$, we have to first normalize the astrophysical model by some DC offset, $F_0$, to yield zero-mean, such that $y(t) = F_0 m_{\mathrm{astro}}(t) + g(t) + \epsilon$, where $m_{\mathrm{astro}}(t)$ is the astrophysical model, $\epsilon$ is the white noise, $g(t) \sim \mathrm{GP}(0,k_{SE})$ (or $k_{M32}$) and $F_0$ is the DC offset. In practice, we profile out the $F_0$ offset, for which the maximum likelihood value occurs at

\begin{align}
\hat{F}_0 &= \frac{ \mathbf{m}^T \mathbf{K}^{-1} \mathbf{y} }{ \mathbf{m}^T \mathbf{K}^{-1} \mathbf{m} }.
\end{align}

Our residuals vector, $\mathbf{r}$, is thus $\mathbf{y} - \hat{F}_0 \mathbf{m}$.

We also experimented with a linear trend (e.g. \expn\ model) with a GP over the top. We ran these models using profiling on the 18 linear terms, as before, but marginalizing the GP parameters. We found that this yielded very similar results to the GP alone. However, a major issue with combining both is that a GP formally violates the assumptions under which the PL method yields accurate posteriors or Bayesian evidences (Moynihan et al. 2025). In Paper I, this was surmountable by simply marginalizing over all 18 linear parameters and abandoning the PL approach, which was practical due to the fact oblate fit only require two extra parameters. But the exomoon model adds seven and these have complex multimodalities that we found made a full marginalization unreliable.

We briefly caution that GPs are inherently very flexible models and they run the risk of absorbing real astrophysical signal - especially when confronted with a single transit as we have here. Thus, some set of real exomoons could be missed by using GPs - in other words, our completeness rate diminishes. However, by demanding exomoons be detectable across a variety of defensible choices for the systematic noise (including these very flexible GPs), the bar is raised and thus our false positive rate decreases. This is the classic trade-off in experimental design. Accordingly, we concede that our approach may not maximize completeness, but any signal that survive will necessarily be highly robust. Given the stakes of a ``first'' confirmed exomoon, we consider that tradeoff not only acceptable but imperative.

\section{Fitting}
\label{sec:fitting}

\subsection{Astrophysical Model}
\label{sub:astromodel}

In this work, we consider two astrophysical models, $f(t)$. The first is a canonical Mandel-Agol planetary transit \citep{mandel:2002} on a circular orbit with quadratic limb darkening\footnote{
As an alternative, we tried using freely fitted four-parameter non-linear limb darkening for our planet-only fits and found the improvement to be marginal at just $\Delta\chi^2=0.3$ and thus we conclude the quadratic parameterization is sufficient for this data set.
} characterized as $q_1$ and $q_2$ following \citep{q1q2:2013}. The model also accounts for the finite integration time of our binned light curve (5\,minutes) via the numerical resampling approach described in \citet{binning:2010} using $N_{\mathrm{resam}} = 10$. This model includes seven free parameters: the planet-to-star ratio-of-radii, $p$; the impact parameter, $b$; the mean stellar density, $\rho_{\star}$; the time of transit minimum, $\tau$; the orbital period, $P$; and the two aforementioned limb darkening coefficients. Uniform priors are adopted for all with two exceptions, $P$ and $\rho_{\star}$.

For $\rho_{\star}$, one might naively turn to spectroscopy-derived constraints to impose an informative prior. However, since our model imposes a circular orbit, $\rho_{\star}$ is really more a pseudo-density conditioned upon that specific assumption \citep{burke:2008,gilbert:2022}. We thus instead derive a prior by re-fitting the previous \kepler-only transits presented in \citet{kipping:2016} under the same model assumptions. This yields a prior of $\rho_{\star} \sim \mathcal{N}(2819.5,114.6)$\,kg\,m$^{-3}$ as an informative prior here.

For the orbital period, we do the same except as fold in the partial \textit{Spitzer} transit reported in \citet{dalba:2019}, giving us $P \sim \mathcal{N}(1071.23325,0.00055)$\,days, in close agreement with the value found by \citet{dalba:2019}.

The second model we consider is the planet+moon model described in \citet{luna:2011}, known as \luna. This analytic \fortran\ implementation remains the fastest modeling code to our knowledge (e.g. see \citealt{kipping:2025}) and fully handles the complex geometry of possible szygy events\footnote{
This is when the projected disks of the star, planet and moon all overlap.
}. Our model does not explicitly include multiple moons ab-initio, although these can be added as required. Should strong evidence for a moon be obtained, but it is a signal driven by multiple moons, the derived parameters will be biased as a result \citep{teachey:2024} - but we refrain from speculating on multiple moons until at least some positive evidence is found to begin with.

The planet-only model, $\mathcal{M}_P$, can be understood as a nested model of the planet+moon model, $\mathcal{M}_S$. The latter appends seven new parameters: the reciprocal of the satellite's orbital period, $1/P_S$ (motivated by \citealt{corridor:2021}); mean density of the planet, $\rho_P$ (to impose physical bounds); cosine of the satellite's orbital inclination relative to the planetary planet, $\cos i_S$ (to ensure isotropic prior); longitude of the ascending node of the satellite orbit, $\Omega_S$; mean anomaly of the exomoon, $\phi_S$; radius of the satellite relative to the planet, $R_{SP}$; and, mass of the satellite relative to the planet, $M_{SP}$. We adopt uniform priors for all with the exception of $\rho_P$ for which we adopt an informative prior leveraging the mass constraint from \citet{chachan:2022}, such that $\rho_P \sim \mathcal{N}(1840,290)$\,kg\,m$^{-3}$. Table~\ref{tab:priors} provides a summary of the parameters and their priors.

\begin{deluxetable*}{llc}
\tablecaption{Parameters sampled over in the light curve fits of this work, along with their associated priors, where $\mathcal{U}$ denotes a uniform distribution, $\mathcal{J}$ denotes a log-uniform distribution and $\mathcal{N}$ denotes a normal distribution.
\label{tab:priors}}
\tabletypesize{\scriptsize}
\tablehead{
\colhead{Parameter} & \colhead{Definition} & \colhead{Prior}
}
\startdata
\cutinhead{\custom}
$R_P/R_{\star}$                & Planet-to-star radius ratio & $\mathcal{U}[0,0.25]$ \\
$\rho_{\star}$\,[kg\,m$^{-3}$] & Mean stellar density & $\mathcal{N}[2819,115]$ \\
$b$                            & Impact parameter & $\mathcal{U}[0,1.25]$ \\
$P_P$\,[days]                  & Planet's orbital period & $\mathcal{N}[1071.23325,0.00055]$ \\
$\tau - $60608\,MJD            & Planet's mid-transit time & $\mathcal{U}[0.4519,1.4519]$\\
$q_1$                          & Limb darkening coefficient & $\mathcal{U}[0,1]$ \\
$q_2$                          & Limb darkening coefficient &$\mathcal{U}[0,1]$ \\
$1/P_S$\,[days$^{-1}$]         & Satellite's orbital period & $\mathcal{U}[0.029,3.339]$ \\
$\rho_P$\,[kg\,m$^{-3}$]       & Mean planetary density & $\mathcal{N}[1840,290]$ \\
$\phi_S$\,[rads]               & Satellite's phase & $\mathcal{U}[0,2\pi]$ \\
$\cos(i_S)$                    & Satellite's inclination & $\mathcal{U}[-1,1]$ \\
$\Omega_S$\,[rads]             & Satellite's longitude of the ascending node & $\mathcal{U}[0,2\pi]$ \\
$M_S/M_P$                      & Satellite-to-planet mass ratio & $\mathcal{U}[0,1]$ \\
$R_S/R_P$                      & Satellite-to-planet radius ratio & $\mathcal{U}[0,1]$ \\
$\sigma_{GP}$                  & GP amplitude & $\mathcal{J}[400,40000]$ \\
$l_{GP}$\,[days]               & GP length scale & $\mathcal{J}[0.0035,1.0]$ \\
\enddata
\end{deluxetable*}

\subsection{Fitting Algorithm}
\label{sub:fits}

The choice of how to sample the target distribution, the sampling algorithm, is important in our problem since the planet+moon problem is highly multimodal \citep{luna:2011}. The outside-in, multimodal sampling approach of \multi\ \citep{feroz:2008,feroz:2009} has been showed to work well for this problem (e.g. see \citealt{hek4} and \citealt{hek5}) and thus we adopt this in what follows. For all of our fits, we use 4000 live points and an enlargment factor of 0.1, following the guidance of \citet{feroz:2009} for reliable evidences, and supported by injection-recovery tests for this problem (e.g. see \citep{luna:2011}; \citealt{hek3}).

We note that \citet{heller:2024} previously questioned the reliability of \multi\ and instead used the \ultra\ code \citep{ultra:2021}. \ultra\ includes multiple sampling methods, including the ellipsoidal sampling method employed by \multi, and the user is required to carefully select the appropriate sampling scheme and associated tuning parameters. As demonstrated in \citet{kipping:2025}, \citet{heller:2024} actually missed the global likelihood in their \ultra\ fits as a result of running in step sampling mode presumably without sufficient experimentation to tune the associated input terms. This highlights the importance of careful use of such codes.

\subsection{A Matrix of Results}

In Section~\ref{sec:observations}, we described our three independent reductions of the data: i) \custom\ (this work), ii) \jedi\ \citep{jedi:2022}, and, iii) \kat\ (in prep.). In Section~\ref{sec:trendmodels}, we described three models for the exposure long trend - the dominant systematic observed in our time series; i) quadratic (\qudr) ii) exponential + linear (\expn), iii) a zero-mean Gaussian process with a squared-exponential kernel (\gp), and, iv) a zero-mean Gaussian process with a Matern-3/2 kernel (\matern). We thus have a 3x4 matrix of results to consider, where for each one we regress models $\mathcal{M}_P$ and $\mathcal{M}_S$ using \multi. For each, we compare the Bayes factor (using the marginal likelihoods estimated by \multi) in support of $\mathcal{M}_S$. As an additional check, we use the marginalized $R_{SP}$ posterior to infer the Bayes factor independently using the Savage-Dickey theorem \citep{dickey:1971}. However, we caution the Bayes factor obtained this way will not be identical since it considers model intersection occurring at $R_{SP}\to0$ whereas strictly we require $M_{SP}\to0$ as well. However, any real detection should produce a convincing posterior peak off-zero in $R_{SP}$, which is the Savage-Dickey ratio tracks. Our detection criterion for a positive-signal is that both Bayes factors exceed 10 (``strong evidence''). The results are summarized in Table~\ref{tab:matrix}.

\begin{deluxetable*}{lccc}
\tablecaption{Summary of key metrics from our matrix of models/reductions. The three columns denote our different reductions, and the four rows denote our different trend models. In each box, the upper number is the 1\,$\sigma$ credible interval on $R_{SP}$, and the lower numbers are the log base ten Bayes factors in favor of the moon model versus the planet model - the first being the formal \multi\ Bayes factor and the latter being that from the Savage-Dickey ratio. If both exceed 1 (Bayes factor of 10), the $R_{SP}$ value is colored teal to denote a positive detection.
\label{tab:matrix}}
\tablehead{
\colhead{} & \colhead{\custom} & \colhead{\jedi} & \colhead{\kat}
}
\startdata
\qudr  & \vstack{\textcolor{teal}{$0.0815_{-0.0064}^{+0.0060}$}}{$\{3.50,>2\}$} & \vstack{\textcolor{teal}{$0.0960_{-0.0038}^{+0.0036}$}}{$\{26.96,>2\}$} & \vstack{\textcolor{teal}{$0.0930_{-0.0064}^{+0.0059}$}}{$\{6.46,>2\}$} \\
\hline
\expn  & \vstack{$0.0683_{-0.0065}^{+0.0074}$}{$\{1.08,0.27\}$} & \vstack{\textcolor{teal}{$0.0861_{-0.0045}^{+0.0043}$}}{$\{15.79,>2\}$} & \vstack{\textcolor{teal}{$0.1067_{-0.0159}^{+0.0092}$}}{$\{3.81,>2\}$} \\
\hline
\gp    & \vstack{$0.056_{-0.029}^{+0.043}$}{$\{-1.67,-0.70\}$} & \vstack{\textcolor{teal}{$0.160_{-0.016}^{+0.015}$}}{$\{2.07,>2\}$} & \vstack{$0.0597_{-0.0062}^{+0.0056}$}{$\{0.62,1.82\}$} \\
\hline
\matern& \vstack{$0.081_{-0.057}^{+0.023}$}{$\{-2.20,-0.74\}$} & \vstack{\textcolor{teal}{$0.179_{-0.018}^{+0.016}$}}{$\{1.17,>2\}$} & \vstack{$0.0647_{-0.0076}^{+0.0063}$}{$\{-0.84,0.82\}$} \\
\enddata
\end{deluxetable*}

\begin{figure*}
\centering
\includegraphics[width=15.5 cm]{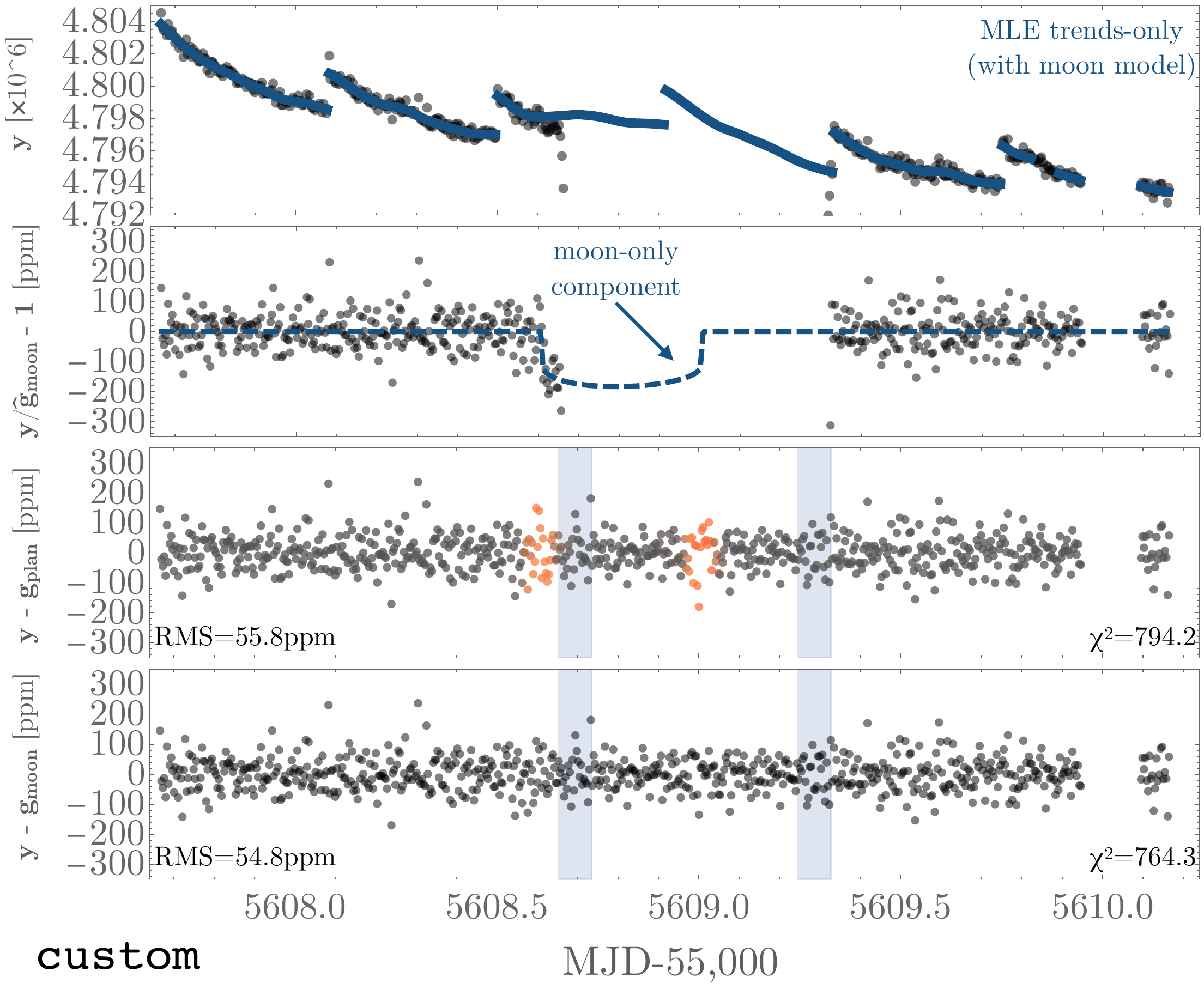}
\caption{\emph{
Overview of the \custom\ \matern\ results.
Top panel zooms in on the light curve, $\boldsymbol{y}$ (dark points), at the scale of the exposure long trends, with the maximum likelihood estimated (MLE) of the trends-only ($\mathbf{\hat{g}}$) shown in blue (derived with the planet+moon model, $\mathcal{M}_S$). 
Second panel shows the data normalized by the trends model (gray points) with the moon-only component from our MLE solution in blue-dashed.
Third panel shows the residuals from a planet-only model ($\mathcal{M}_P$), with a color-scaling from gray to red in proportion to their relative contribution to the $\chi^2$ improvement of the planet-moon model ($\mathcal{M}_S$). We also highlight the planetary ingress/egress with shading.
Bottom panel shows the residuals of the planet-moon model ($\mathcal{M}_S$).
}
} 
\label{fig:m32custom}
\end{figure*}

\begin{figure*}
\centering
\includegraphics[width=15.5 cm]{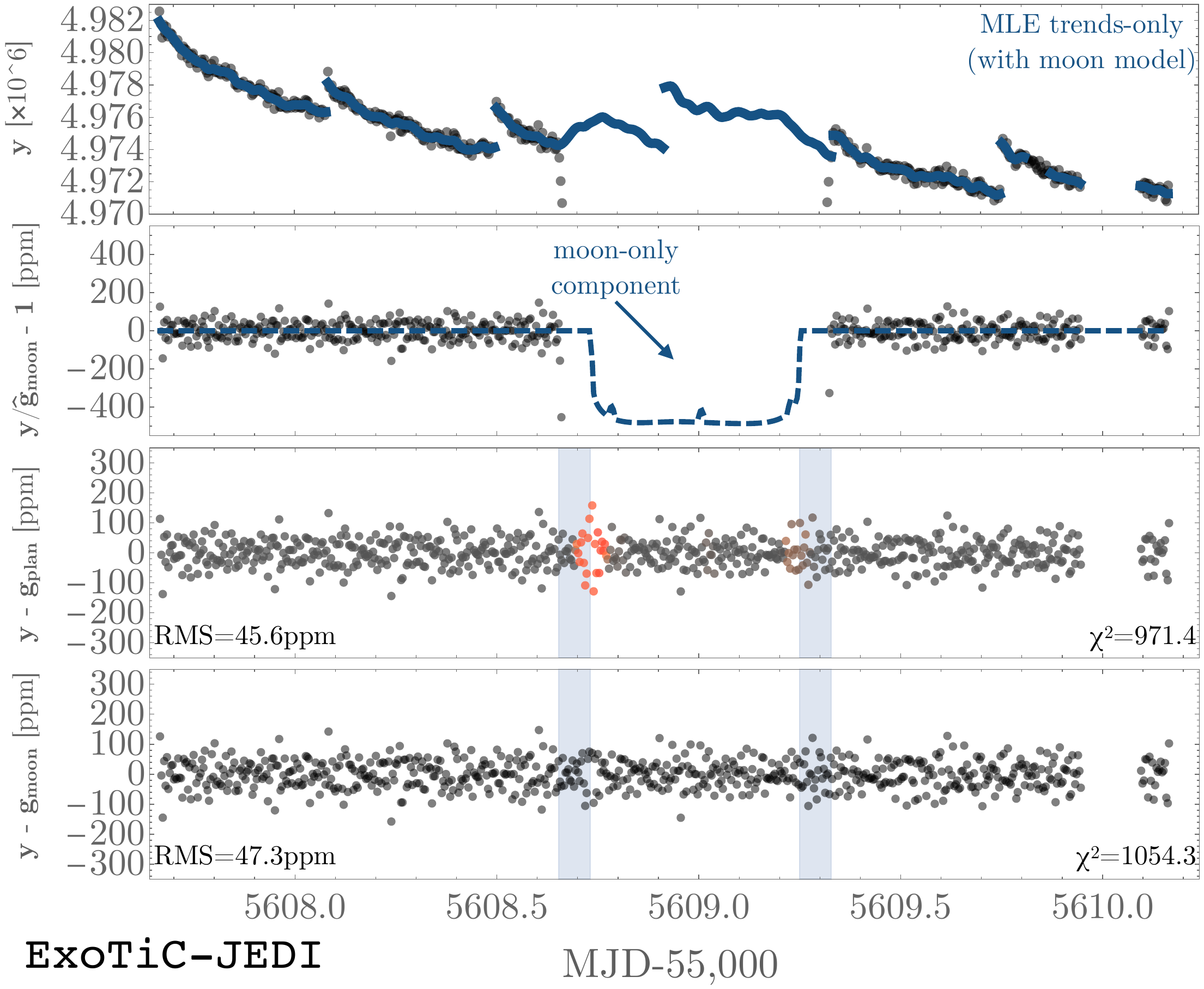}
\caption{\emph{
Overview of the \jedi\ \matern\ results. Same as style as Figure~\ref{fig:m32custom}.
}
} 
\label{fig:m32jedi}
\end{figure*}

\begin{figure*}
\centering
\includegraphics[width=15.5 cm]{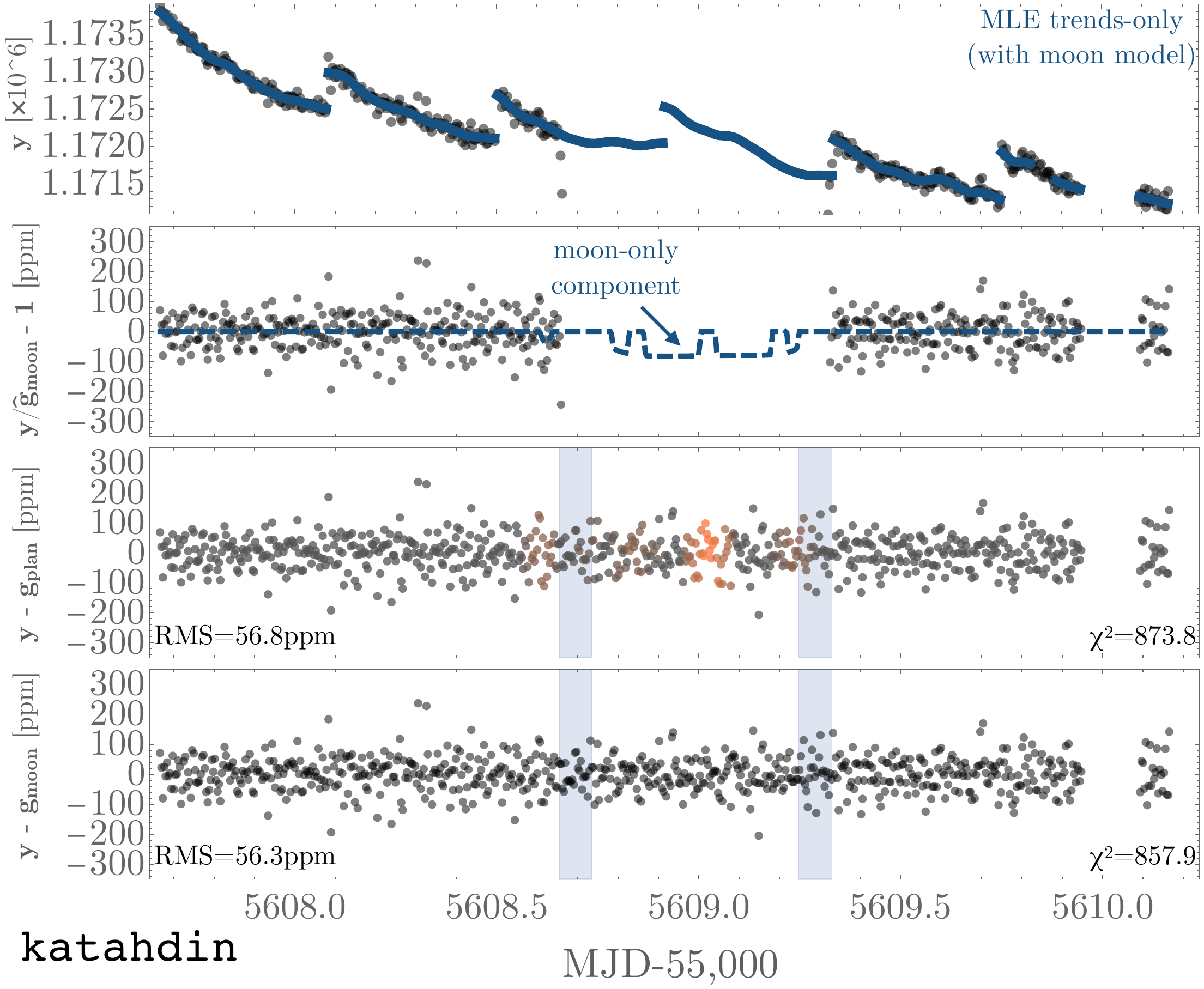}
\caption{\emph{
Overview of the \kat\ \matern\ results. Same as style as Figure~\ref{fig:m32custom}.
}
} 
\label{fig:m32katahdin}
\end{figure*}

At the broadest level, we found that 7 out of our 12 fits were classified as positive detections. Amongst these, and even for the non-detections, the recovered signal is almost always that of moon-like dip almost precisely on-top of the planetary transit with one or more syzgy events (see second panel of Figures~\ref{fig:m32custom}-\ref{fig:m32katahdin} \& \ref{fig:expcustom}-\ref{fig:expkatahdin}), with $R_{SP}$ spanning 0.06 to 0.18.

Breaking this down a little deeper, it is notable that 5 of these detections occur for \qudr\ and \expn\ models (indeed every \qudr\ model yields a detection), whereas amongst the remaining six GP models only two detections were found. Of these two GP detections, both were attached to the \jedi\ reduction, yielding the two largest inferred moon radii out of the matrix of 12: $R_{SP} \sim 0.17$. As these results hint at, we found the \qudr\ and \expn\ produced broadly similar results to each other, and the two GPs models were also self-similar. As a result, and for the sake of conciseness, we limit many of our figures to just the \expn\ and \matern\ models in what follows.

\section{Analysis}
\label{sec:analysis}

\subsection{Revised Planetary Parameters}

Before we discuss the moon fits in detail, the primary focus of this paper, we briefly present revised transit parameters for Kepler-167e from our planet-only fits. Table~\ref{tab:plan} summarizes the one-sigma credible intervals on the planetary transit parameters.

\begin{deluxetable*}{lcccc}
\tablecaption{Summary of revised one-sigma credible intervals for Kepler-167e with new JWST NIRSpec data, grouped by reduction. We omit orbital period since it is held by a strongly informative prior.\label{tab:plan}}
\tabletypesize{\scriptsize}
\tablehead{
\colhead{Parameter} & \colhead{\qudr} & \colhead{\expn} & \colhead{\gp} & \colhead{\matern}
}
\startdata
\cutinhead{\custom}
$R_P/R_{\star}$                & $0.12531_{-0.00021}^{+0.00022}$ & $0.12537_{-0.00035}^{+0.00035}$ & $0.12620_{-0.00064}^{+0.00064}$ & $0.12594_{-0.0006}^{+0.0006}$ \\
$b$                            & $0.2893_{-0.0058}^{+0.0056}$    & $0.2879_{-0.0062}^{+0.0060}$    & $0.2802_{-0.0088}^{+0.0089}$    & $0.2838_{-0.0092}^{+0.0088}$   \\
$\rho_{\star}$\,[kg\,m$^{-3}$] & $2756_{-14}^{+14}$              & $2757_{-14}^{+15}$              & $2776_{-22}^{+21}$              & $2765_{-21}^{+21}$             \\
$\tau - $60608\,MJD        & $0.990854_{-0.000067}^{+0.000068}$ & $0.990861_{-0.000070}^{+0.000067}$ & $0.990729_{-0.000086}^{+0.000083}$ & $0.990720_{-0.00013}^{+0.00012}$ \\
$q_1$                          & $0.1589_{-0.0061}^{+0.0063}$    & $0.1633_{-0.0061}^{+0.0066}$    & $0.169_{-0.011}^{+0.011}$       & $0.168_{-0.013}^{+0.015}$      \\
$q_2$                          & $0.385_{-0.023}^{+0.023}$       & $0.366_{-0.030}^{+0.031}$       & $0.440_{-0.069}^{+0.069}$       & $0.402_{-0.067}^{+0.066}$      \\
\tableline
\cutinhead{\jedi}
$R_P/R_{\star}$                & $0.12342_{-0.00017}^{+0.00016}$ & $0.12328_{-0.00026}^{+0.00027}$ & $0.12423_{-0.00059}^{+0.00058}$ & $0.12448_{-0.00070}^{+0.00068}$ \\
$b$                            & $0.2926_{-0.0047}^{+0.0045}$    & $0.2887_{-0.0049}^{+0.0049}$    & $0.274_{-0.012}^{+0.012}$       & $0.284_{-0.019}^{+0.016}$      \\
$\rho_{\star}$\,[kg\,m$^{-3}$] & $2727_{-11}^{+11}$              & $2733_{-12}^{+12}$              & $2743_{-28}^{+27}$              & $2726_{-38}^{+40}$             \\
$\tau - $60608\,MJD         & $0.990930_{-0.000055}^{+0.000052}$ & $0.990937_{-0.000055}^{+0.000056}$ & $0.990480_{-0.00027}^{+0.00027}$ & $0.990540_{-0.00037}^{+0.00036}$ \\
$q_1$                          & $0.1768_{-0.0054}^{+0.0054}$    & $0.1862_{-0.0056}^{+0.0057}$    & $0.309_{-0.038}^{+0.041}$       & $0.258_{-0.044}^{+0.053}$      \\
$q_2$                          & $0.316_{-0.017}^{+0.017}$       & $0.273_{-0.023}^{+0.023}$       & $0.088_{-0.056}^{+0.071}$       & $0.162_{-0.086}^{+0.097}$      \\
\tableline
\cutinhead{\kat}
$R_P/R_{\star}$                & $0.12583_{-0.00021}^{+0.00021}$ & $0.12585_{-0.00034}^{+0.00033}$ & $0.12699_{-0.00063}^{+0.00062}$ & $0.12662_{-0.00067}^{+0.00068}$ \\
$b$                            & $0.2879_{-0.0056}^{+0.0056}$    & $0.2852_{-0.0060}^{+0.0059}$    & $0.2785_{-0.0088}^{+0.0087}$    & $0.2855_{-0.0107}^{+0.0099}$   \\
$\rho_{\star}$\,[kg\,m$^{-3}$] & $2758_{-13}^{+13}$              & $2762_{-14}^{+14}$              & $2780_{-21}^{+21}$              & $2761_{-24}^{+25}$             \\
$\tau - $60608\,MJD         & $0.990920_{-0.000065}^{+0.000067}$ & $0.990922_{-0.000070}^{+0.000069}$ & $0.990837_{-0.000082}^{+0.000082}$ & $0.990790_{-0.00017}^{+0.00016}$ \\
$q_1$                          & $0.1577_{-0.0059}^{+0.0061}$    & $0.1635_{-0.0060}^{+0.0062}$    & $0.164_{-0.010}^{+0.011}$       & $0.164_{-0.018}^{+0.020}$      \\
$q_2$                          & $0.370_{-0.022}^{+0.022}$       & $0.346_{-0.029}^{+0.029}$       & $0.449_{-0.068}^{+0.068}$       & $0.388_{-0.090}^{+0.088}$      \\
\enddata
\end{deluxetable*}

Perhaps the most useful set of numbers to the community is an average over all 12 models. Fr simplicity, we give each model equal weight and combine to yield final credible intervals of
$R_P/R_{\star} = 0.12550_{-0.00193}^{+0.00099}$,
$b=0.2860_{-0.0107}^{+0.0078}$,
$\rho_{\star} = 2755_{-27}^{+24}$\,kg\,m$^{-3}$,
$\tau - $60608\,MJD$=0.99085_{-0.00021}^{+0.00011}$,
$q_1 = 0.169_{-0.013}^{+0.039}$ \&
$q_2 = 0.354_{-0.115}^{+0.076}$. Even after averaging, it is noteworthy that the pseudo-density is constrained to within one percent by JWST.

\begin{figure*}
\centering
\includegraphics[width=15.5 cm]{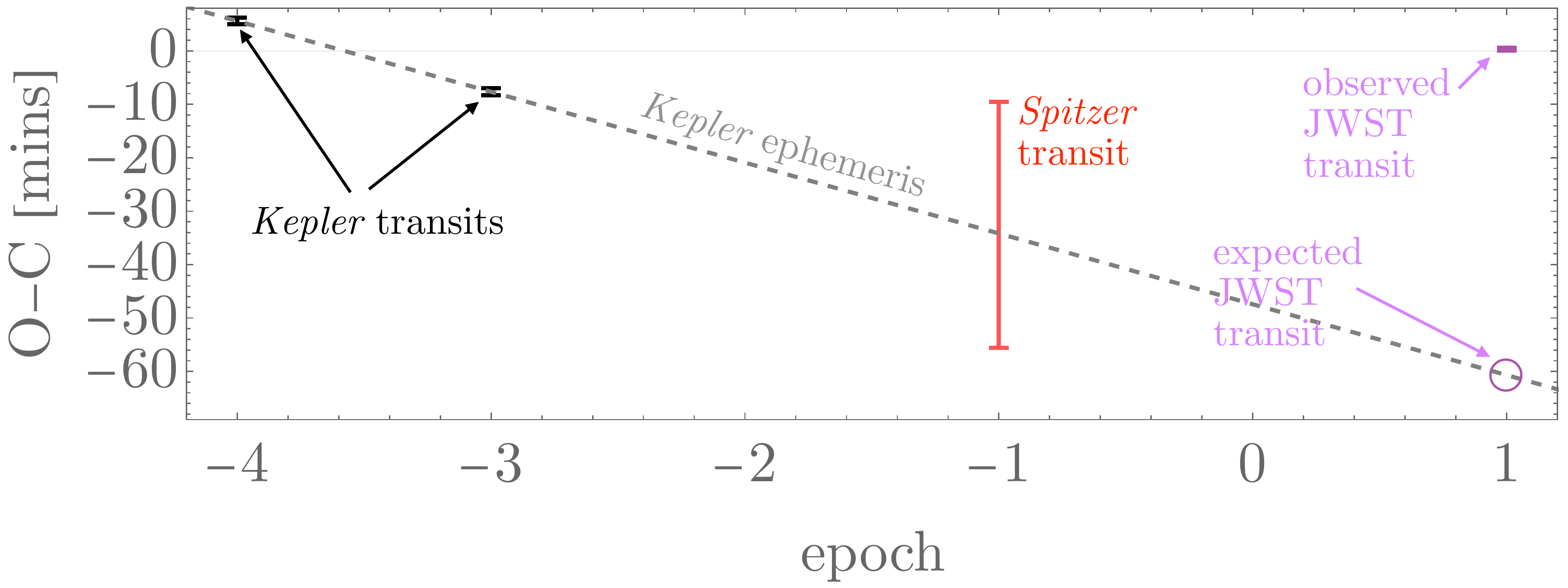}
\caption{\emph{
The four transit times observed to date of Kepler-167e. The JWST transit appears to have transited approximately an hour later than expected, given the \kepler\ transit times. The new ephemeris implies residual TTVs of order 10\,minutes.
}
} 
\label{fig:TTVs}
\end{figure*}

It also notable that the JWST transit occurred just over an hour later than expected by comparison to the \citet{kipping:2016} and \citet{dalba:2019} ephemerides (see Figure~\ref{fig:TTVs}). Re-fitting a new ephemeris, we find $\tau = 2459538.2491\pm0.0019$\,BJD and $P=1071.24151\pm0.00087$\,days, although clearly there are TTVs on top of this. Including the \textit{Spitzer} transit time, the TTVs would have a standard deviation of 16.9\,minutes. Ignoring this point, the TTV standard deviation would be 6.7\,minutes. More observations are clearly needed to verify this deviation, which may indicate a second gas giant in the outskirts of the Kepler-167 system.

\subsection{\jedi\ GP outlier}
\label{sub:jedioutlier}

The \jedi\ GP results represent the greatest outlier amongst our grid of results. First, the inferred moon radius ($R_{SP} \sim 0.17$) is much higher than the other cases ($R_{SP} \sim 0.08$); see Table~\ref{tab:matrix}. More importantly though, inspection of the \gp\ and \matern\ moon fits (see top panel of Figure~\ref{fig:m32jedi}) shows how the trend model introduces a pair of large step functions almost exactly on top of the planetary transit, which the moon then accounts for. We do not see analogous behavior in our other GP model-reduction combinations (e.g. see Figures~\ref{fig:m32custom} and \ref{fig:m32katahdin}).

As the top panels of Figures~\ref{fig:m32custom}-\ref{fig:m32katahdin} show, this behavior is also not observed in the four out-of-transit exposures (nor indeed other JWST programs). Together, these strongly suggest the GP \jedi\ moon solutions are spurious. We are unaware of a moon model ever contriving with a trend model to act so strangely in previous moon transit searches, and we speculate that this may be a product of the fact that the trend timescale is very similar to the planet (and thus moon) transit duration. However, is unclear why the \jedi\ reduction would be more susceptible to this effect than the other two considered.

Although there is already sufficient reason to consider the GP \jedi\ moon solutions spurious, we will make use of two other useful tests in the context of a clear false-positive. First, the \jedi\ \matern\ solution is highly chromatic. This was found by first splitting the spectrum into a red/blue channel of roughly equal flux and then repeating our fits. The red channel yields a strong detection of $R_{SP}=0.164_{-0.023}^{+0.021}$, broadly consistent with that from the white light curve, whereas the blue channel yields a non-detection of $R_{SP}=0.052_{-0.022}^{+0.011}$. Of course, a moon should not be highly chromatic like this.

The second test is to repeat the white light curve fits allowing for negative-radius moons, which manifest as inverted transits within \luna. Whereas the \custom\ reduction yield non-detections of inverted moons with a GP model, the \jedi\ model again yields infers an enormous negative moon each time, $R_{SP} = -0.128_{-0.012}^{+0.012}$ (\gp) and $R_{SP} = -0.135_{-0.014}^{+0.014}$ (\matern). This, again, is not the expected outcome of a real signal.

\subsection{Signal Location Analysis}
\label{sub:siglocation}

\begin{figure*}
\centering
\includegraphics[width=15.5 cm]{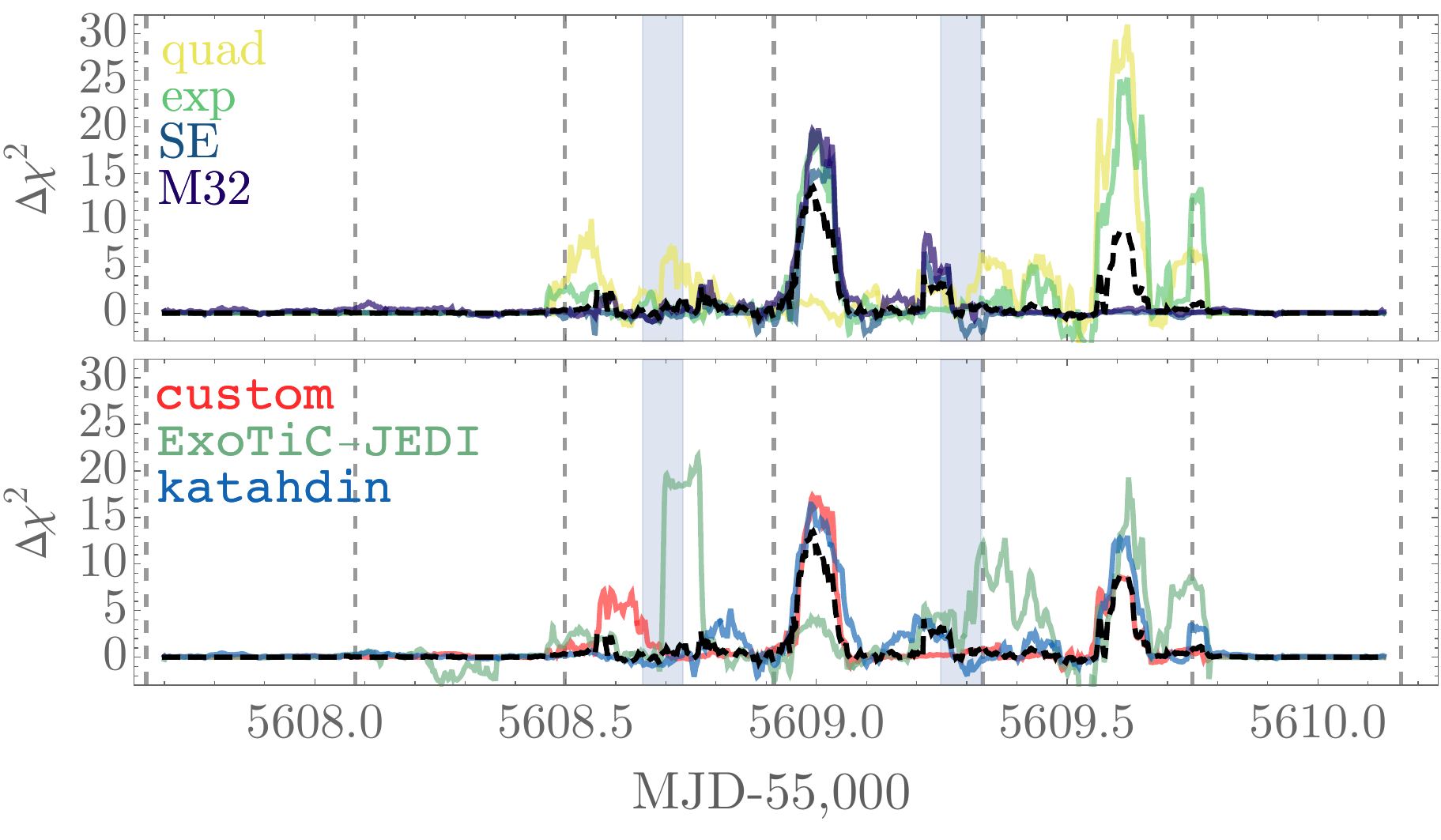}
\caption{\emph{
Location of the moon signals. Each line represents a moving median of the $\Delta\chi^2$ improvement of the moon model over the planet model, over a 100-minute window. This allows one to see which parts of the time series most contribute to putative detections. The top panel shows the median curves of each trend model (averaged over reductions). The bottom panel shows the median curves of each reduction (averaged over trend models). In both, the dashed black line is the ensemble median, the shaded regions are the planetary ingress/egress and the vertical dashed lines are the exposure transition points.
}
} 
\label{fig:siglocation}
\end{figure*}

In Figure~\ref{fig:siglocation}, we used a 100-minute moving median of the $\Delta\chi^2$ between models $\mathcal{M}_S$ and $\mathcal{M}_P$, in order to track which temporal regions of the light curve most greatly contribute to the recovered signals. The black dashed line is averaged over all 12 models and reveals two dominant regions in time: a syzygy-like feature located near the mid-point of the planetary transit, and a dip-like feature located post-transit. The former dominates in most models and manifests as szygy in our models, which is highly ambiguous with a spot crossing event \citep{rabus:2009,ojeda:2011,beky:2014}. The latter is thus potentially more promising as a clear exomoon signature, since a spot crossing cannot explain such a feature \citep{teachey:2020}.

We note that the \jedi\ reduction diverges somewhat in signal location from the other two, as evident from the lower panel of Figure~\ref{fig:siglocation}. Whilst it picks up the same two dominant features as the others (albeit attenuating the syzygy-like feature), it adds additional features approximately where the planetary ingress and egress occur.

Across the different trend models, we find that \expn, \gp\ and \matern\ all support the mid-transit szygy and the only countering voice is that of the \qudr\ model, giving a 3:1 split. For the second out-of-transit feature, the vote is 2:2, with the GP models both removing this feature but the linear models supporting it. Since we take a median of four models, this explains the attenuated ensemble signal at this location in black-dashed. Given that this feature is not favored by most of our models, and is not recovered by the agnostic GP models, we consider it to be spurious in what follows. Thus, the only feature that appears to be likely real is the mid-transit syzygy-like event.

\subsection{Masking the Syzygy-like Event}

The \custom\ and \kat\ reductions, when using a GP trend model, formally do not imply a moon detection (see Table~\ref{tab:matrix}). Despite this, the MLE moon fits still trace out the syzygy-like feature present near the mid-point of the planetary transit. Although the fits are not statistically significant detections, we hypothesized that their support for the moon model would be even more diminished if we were to mask the syzygy-like event.

We thus masked data from MJD 60608.92 to 60609.07 and repeated our two GP models for all three reductions. We found that the Savage-Dickey ratio in support of hypothesis $\mathcal{M}_S$ decreased in all six cases.

From Table~\ref{tab:matrix}, the \custom\ reduction yields the most conservative set of results in favor of a moon, and similarly the GP models are consistently more conservative than the linear models. We thus treat the \custom\ with \matern\ model as our favored final model amongst the six masked-syzygy runs. We note that, by design, \custom\ is highly conservative and this is reflected in its relative performance. Further, \matern\ consistently yields more conservative Bayes factors than \gp, perhaps as a result of relaxing the need for smooth functions.

\section{Discussion}
\label{sec:discussion}

\subsection{Sensitivity Estimate}

From our preferred model of \matern\ with \custom\ masking the szygy-like feature, we thus find a posterior of $R_{SP} = 0.038_{-0.025}^{0.039}$ [$<0.094$ to 95\%], which equates to $R_S = 0.37_{-0.24}^{+0.39}$\,$R_{\oplus}$ [$<0.95$\,$R_{\oplus}$ to 95\%] or $R_S/R_{\star})^2 = 22_{-20}^{+70}$\,ppm [$<139$\,ppm to 95\%]. This sensitivity is certainly below what was anticipated in planning these observations, where injections of a Ganymede (0.41\,$R_{\oplus}$) into simulated light curves were recovered in $83$\% of realizations to Bayes factors $>10$. We note that these pre-observation simulations yielded a more conservative precision than what was suggested by \pandexo. We used \pandeia \citep{pontoppidan:2016}, the engine behind \pandexo and the official \jwst\ exposure time calculator, to extract the expected wavelength-dependent photometric precisions. We then discarded all wavelength bins that were flagged as partially saturated on the last group. This is distinct from the default \pandexo routine, which assumes that information can still be extracted from partially saturated pixels via fits to only the first few groups. By entirely discarding all columns that exhibited a chance of partial saturation, we did not consider the very highest signal-to-noise wavelengths, and thus ended at an expected precision nearly 28\% worse than pandexo’s prediction (550 versus 440\,ppm per 1.6\,s integration). Regardless, our exomoon sensitivity from the real observations is certainly lower than our estimates. We consider that there are three primary reasons for this mismatch, which are relevant lessons for future searches.

First, we had anticipated a white light curve precision of 39\,ppm per 5\,minute bin (assuming Gaussian statistics) whereas \custom\ is at 55\,ppm, thus representing 40\% higher noise floor. To some degree, this is a product of \custom's aggressive removal of so many spectral channels in order to fight for Gaussian-like noise statistics. In comparison, \jedi\ is only 22\% above our initial simulations at 47.5\,ppm. But even so, clearly additional noise sources exist beyond what was anticipated.

Second, our moon fits consistently converged to very short period moons ($<$1\,day) and a few planetary radii orbits, placing their transits almost on top of the planet's. We are least sensitive to moons at such small separations, because of the high covariance between the planet and moon morphologies. Indeed, this was anticipated in our proposal where it was noted that we had no sensitivity to an Io or Europa analog, as a result of their small size and tight orbits. In contrast, we were sensitive to 83\% of Ganymede-analogs (15\,$R_P$ from Jupiter) and 53\% of Callisto-analogs (26\,$R_P$). Thus our marginalized posterior of $R_S = 0.37_{-0.24}^{+0.39}$\,$R_{\oplus}$ is not representative of a Ganymede or Callisto analog, but rather a moon even closer in than Io (6\,$R_P$). It may be possible to derive upper limits on the moon radius as a function of semi-major axis from this data set, but that is beyond the scope of this work.

Third, our initial simulations did not include any trends as the studies available at the time of the Cycle 3 call (e.g. \citealt{zafar:2023}; \citealt{alderson_wasp_ers_2023}) indicated that long-term trends could be well-described by simple linear functions. However, this has clearly been the greatest headache in searching for moons in this data set. Our favored approach of a Gaussian process naturally absorbs noise and thus decreases our sensitivity. We consider the problem particularly manifest in this data set as a result of the comparable timescales between that of each exposure and the transit duration.

\subsection{Plausibility of a Spot-Crossing}

The syzygy-like event is highly ambiguous with a spot crossing, but at face value this may be somewhat surprising since Kepler-167 exhibits no detectable rotational modulations nor spot crossings in the \kepler\ data set (although only two transits were captured). We refrain from detailed spot modeling here, given the fact we have just a single event and no measurement of a stellar rotation period. Nevertheless, it is worthwhile to consider the feasibility of this feature indeed being caused by spots.

\begin{figure*}
\centering
\includegraphics[width=17.0 cm]{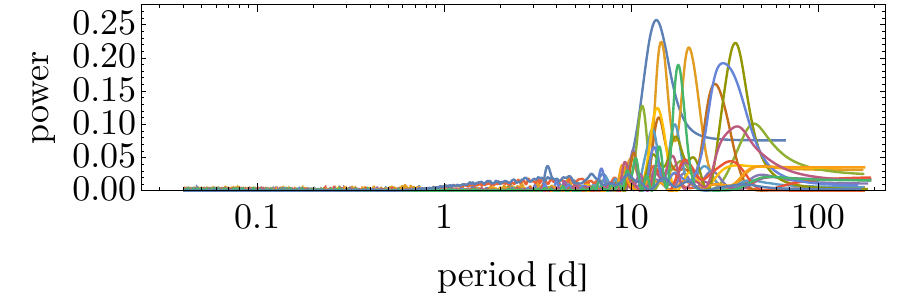}
\caption{\emph{
Lomb-Scargle periodogram of the seventeen available \kepler\ quarters of Kepler-167, revealing no clear coherent rotation period.
}
} 
\label{fig:rotation}
\end{figure*}

We ran a Lomb-Scargle periodogram through every the 17 available \kepler\ quarters of Kepler-167, using the PDC data. We find no coherent rotation period across each quarter (top panel of Figure~\ref{fig:rotation}), with a range of favored periods between 6.1-37.0\,days. For $P<50$\,days (to avoid contamination by residual trends in each 90\,day quarter), the maximum sinusoidal signal recovered (lower panel of Figure~\ref{fig:rotation}) had an amplitude as small as 62\,ppm (quarter 13) and as high as 338\,ppm (quarter 2). We thus assume that the maximum plausible \kepler-band effective spot induces a 338\,ppm variation.

With a single bandpass, there is no way to break the degeneracy between spot size and contrast. However, Kepler-167 ($T_{\mathrm{eff}}=4830$\,K, $M_{\star} = 0.78$\,$M_{\odot}$; \citealt{chachan:2022}) has similar fundamental parameters to HAT-P-11 $T_{\mathrm{eff}}=4780$\,K, $M_{\star} = 0.81$\,$M_{\odot}$; \citealt{kipping:2016}), for which the spots are measured to be 280\,K cooler than photosphere \citep{schutte:2023} and thus we assume the same here. Integrating over the \kepler\ bandpass, a 350\,ppm spot of this contrast would need to be $1.5$\,$R_{\oplus}$. Such a spot could be easily fully eclipsed by the much larger Kepler-167e ($10.2$\,$R_{\oplus}$; \citealt{kipping:2016}) and thus a full spot crossing is a-priori more likely than a grazing event. Finally, we need to calculate the amplitude such a spot could produce in the NIRSpec bandpass. Integrating again, we find that such a spot would manifest at the 263\,ppm level. In the moon model, szygy events are caused by fully opaque objects and in this limit 263\,ppm corresponds to a 1.3\,$R_{\oplus}$ object.

Thus, we conclude that spots consistent with the \kepler\ data could produce spot crossings appearing as szygy-like features associated with moons as large as 1.3\,$R_{\oplus}$, or $R_{SP} \sim 0.13$. This is larger than all of the inferred moon sizes in our models, with the exception of the outlier \jedi\ results discussed earlier. Accordingly, we consider that the feature could certainly be a real spot crossing.

\subsection{Lessons for Future Searches}

This study represents the first attempt at searching for transiting exomoons with JWST and thus it is instructive to reflect on lessons for future searches. We highlight three key lessons.

First, at least for this study, the exposure-long trends were the greatest challenge to deal with. To begin with, like the HST visit long trends, the cause and functional form of these trends is not well-understood and thus we are left with little alternative but to essentially guess heuristic models. Without any clear physical guidance from the literature, nor of course a comparison star to train upon, this represents the greatest source of uncertainty: what model to use? Just as using multiple JWST pipelines has become a standard practice in the exoplanet literature, we recommend multiple trend models such as that used here - which clearly produced varying results.

One way to mitigate this effect is perhaps in the planning stages. It seems likely that that our observations were particularly sensitive to modeling choices of the exposure-long trends because each exposure lasted 10\,hours and the planetary transit lasts a comparable 16\,hours. Thus, changing the trend model affects the transit shape in joint-fitting procedures, a challenge to not just exomoon hunting but any other phenomenon at this timescale, such as rings \citep{barnes:2004,rein:2019}. For this transit, it may be advantageous to modify the exposures to either be much greater or less than the transit duration, although this latter choice could plausibly introduce new headaches. Where long exposures are required, another option is to focus on planets with durations much shorter than the expsoure time. Most atmospheric studies are indeed in this regime, which is perhaps why this issue has not been highlighted much previously.

The second lesson is that spot crossings are a major source of false-positive for close-in moons. Of course, this has been anticipated for over a decade (e.g. see \citealt{luna:2011}) but this study arguably demonstrates the clearest example of such an occurrence. Our observations targeted Kepler-167, in part, because it was known to be a relatively quiescent star from \kepler\ data. Yet, JWST is so much more precise that spots could easily be undetectable by \kepler\ yet manifest in these data. Obviously avoiding known active stars is wise, but even doing so we might expect spots to influence the results, as appears likely in this work. One way mitigation strategy is to move even further into the red, leveraging MIRI perhaps, but a proof-of-principle moon search in that mode would greatly inform how useful this would be.

The third lesson is that we have, in effect, a single transit here. Yes, \kepler\ recorded two previous transits \citep{kipping:2016} and \textit{Spitzer} even caught a partial, but far superior capabilities of JWST means these previous transits have negligible constraining power at the precision JWST probes. Thus, we are essentially working with a single transit. In such a scenario, the moon model is enormously malleable; it can explain essentially any slight bump in the data and thus the risk of a type I error is high. A series of transits, especially if sequential, greatly reduces the ability of the moon model to explain anything it wants, for now it must obey a strict Keplerian from epoch-to-epoch. This is not to say that our JWST transit of Kepler-167e, nor indeed similar efforts on other targets, are a waste of time. If we need a series of transits, then that necessarily begins with a single observation. The next transit of Kepler-167e will occur in October 2027, in cycle 6, and we strongly advocate for a second transit - which should greatly help in trimming the noted degeneracies.

\newpage
\appendix
\section{Additional Figures}
\label{app:figs}

\begin{figure*}[b!]
\centering
\includegraphics[width=15.5 cm]{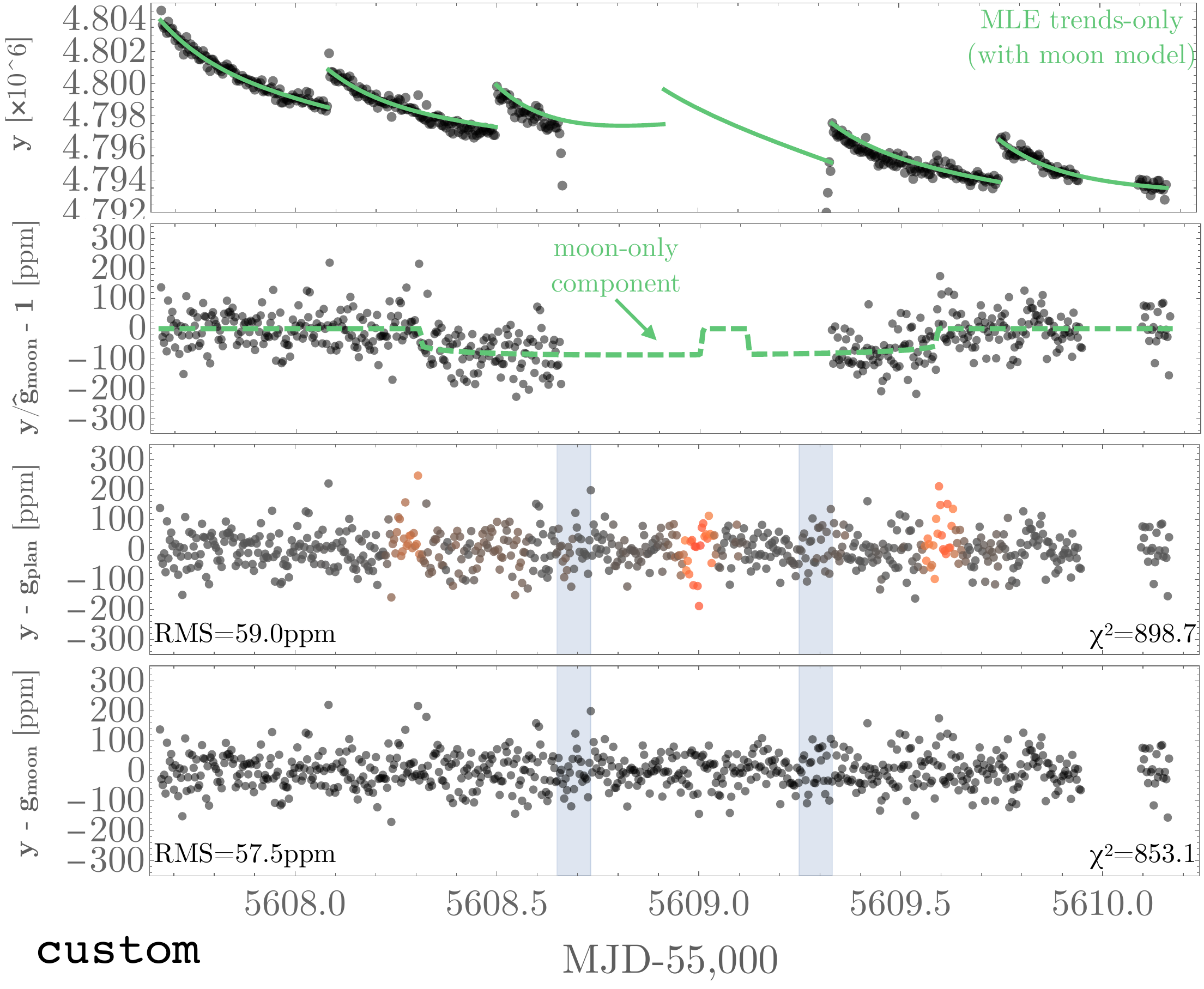}
\caption{\emph{
Overview of the \custom\ \expn\ results. Same as style as Figure~\ref{fig:m32custom}.
}
} 
\label{fig:expcustom}
\end{figure*}

\begin{figure*}[t!]
\centering
\includegraphics[width=15.5 cm]{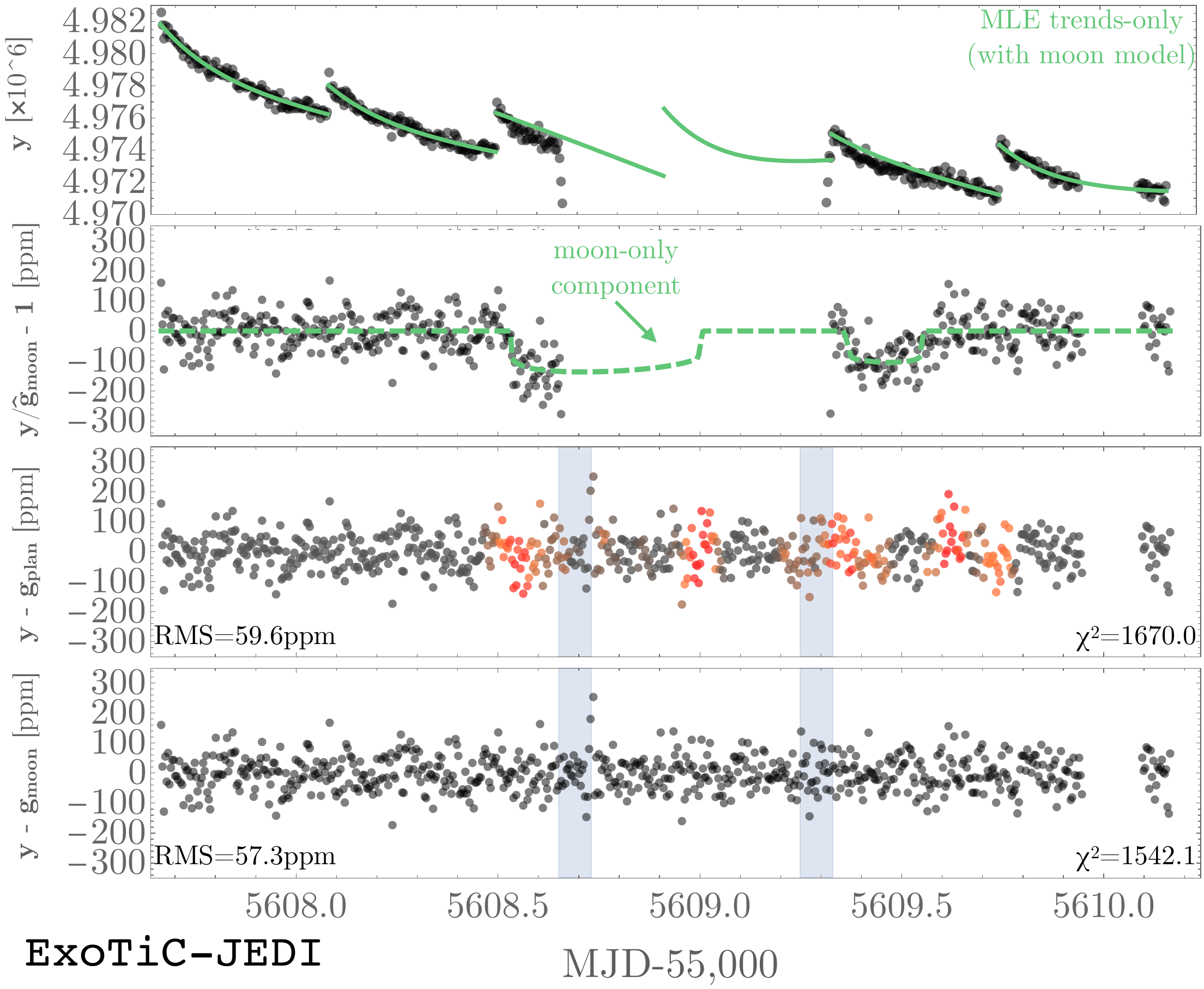}
\caption{\emph{
Overview of the \jedi\ \expn\ results. Same as style as Figure~\ref{fig:m32custom}.
}
} 
\label{fig:expjedi}
\end{figure*}

\begin{figure*}
\centering
\includegraphics[width=15.5 cm]{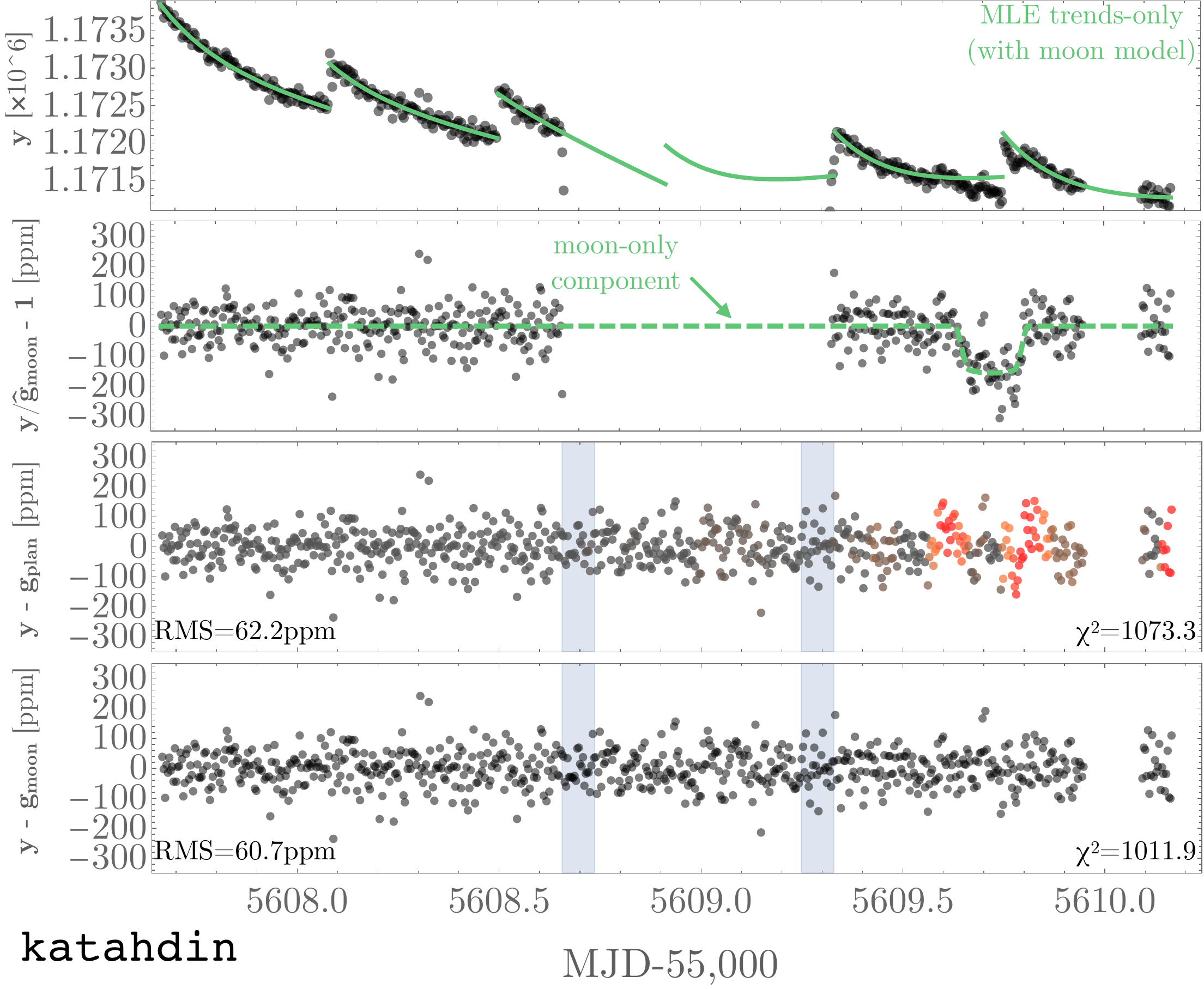}
\caption{\emph{
Overview of the \kat\ \expn\ results. Same as style as Figure~\ref{fig:m32custom}.
}
} 
\label{fig:expkatahdin}
\end{figure*}

In this appendix, we provide supplementary figures which largely mirror Figure~\ref{fig:m32custom} but with different data reductions. Specifically,
Figure~\ref{fig:expcustom} is for \custom\ \expn,
Figure~\ref{fig:expjedi} is for \jedi\ \expn\ and
Figure~\ref{fig:expkatahdin} is for \kat\ \expn.

\section*{Acknowledgements}

This research has made use of the Astrophysics Data System, funded by NASA under Cooperative Agreement 80NSSC21M00561.

We thank the Yale Center for Research Computing for guidance and assistance in the computation run on the Bouchet Cluster.

Resources supporting this work were provided by the NASA High-End Computing (HEC) Program through the NASA Advanced Supercomputing (NAS) Division at Ames Research Center.

This work is based [in part] on observations made with the NASA/ESA/CSA James Webb Space Telescope. The data were obtained from the Mikulski Archive for Space Telescopes at the Space Telescope Science Institute, which is operated by the Association of Universities for Research in Astronomy, Inc., under NASA contract NAS 5-03127 for JWST. These observations are associated with program \#6491. The data can be found in MAST: \dataset[10.17909/e50n-4y96]{http://dx.doi.org/10.17909/e50n-4y96}. 

Special thanks to donors to the Cool Worlds Lab, without whom this kind of research would not be possible:
Douglas Daughaday,
Elena West,
Tristan Zajonc,
Alex de Vaal,
Mark Elliott,
Stephen Lee,
Zachary Danielson,
Chad Souter,
Marcus Gillette,
Jason Rockett,
Tom Donkin,
Andrew Schoen,
Mike Hedlund,
Ryan Provost,
Nicholas De Haan,
Emerson Garland,
Queen Rd Fnd Inc.,
Ieuan Williams,
Axel Nimmerjahn,
Brian Cartmell,
Guillaume Le Saint,
Robin Raszka,
Bas van Gaalen,
Josh Alley,
Drew Aron,
Warren Smith,
Brad Bueche,
Steve Larter,
Marisol Adler \&
Craig Frederick.

\bibliography{manuscript}{}
\bibliographystyle{aasjournalv7_surnames_only}



\end{document}